\documentclass[aps,11pt]{revtex4}
\usepackage[dvips]{graphicx}

\begin{document}
\title{Magnetism and fine electronic structure\\
of UPd$_{2}$Al$_{3}$ and NpPd$_{2}$Al$_{3}$}
\author{R.J. Radwanski}
\homepage{http://www.css-physics.edu.pl}
\email{sfradwan@cyf-kr.edu.pl}
\affiliation{Center for Solid State Physics, S$^{nt}$Filip 5, 31-150 Krakow, Poland,\\
Institute of Physics, Pedagogical University, 30-084 Krakow,
Poland}
\author{Z. Ropka}
\affiliation{Center for Solid State Physics, S$^{nt}$Filip 5,
31-150 Krakow, Poland}

\begin{abstract}
We claim the existence of the $f^{3}$ (U$^{3+}$) configuration in
UPd$_2$Al$_3$. It is in agreement with
inelastic-neutron-scattering (INS) excitations and is consistent
with the trivalent neptunium configuration in NpPd$_2$Al$_3$. We
have derived set of CEF parameters for the U$^{3+}$ state that
reproduces the INS excitations and temperature dependence of the
heat capacity. On basis of the crystal-field theory, extended to
Quantum Atomistic Solid State Theory we argue that the magnetic
moment of the uranium moment amounts at 0 K to 1.7-1.8
$\mu_{B}$.
\end{abstract}
\maketitle
\section{Introduction}
The description of electronic and magnetic properties of
UPd$_2$Al$_3$ is still under hot debate though its exotic
properties has been discovered already more than 10 years ago [1].
The uniqueness of UPd$_2$Al$_3$ relies in the coexistence of the
heavy-fermion (h-f) phenomena and the large magnetic moment of
about 1.5 $\mu _B$ ( T$_N$ =14.3 K) as well as superconductivity
below 2 K. The main point of the debate is related to the
understanding of the role played by $f$ electrons. In the band
calculations these electrons are considered as itinerant [2,3]
whereas surprisingly nice reproduction of many experimental
results can be obtained within the crystal-field (CEF) approach,
i.e. treating f electrons as localized [1,4,5,6]. However, within
the CEF approach there is presently discussion about the
tetravalent [1,4,7,8] or trivalent [5,6] uranium state in
UPd$_2$Al$_3$. As a great progress towards the consensus about the
f electrons in heavy-fermion compounds we take the drastic change
of the physical mind of Fulde and Zwicknagl, who advocating for
years for all f electrons itinerant now propagate a model with a
dual nature of f electrons, two of which are localized [8].

The U$^{3+}$ ion is a Kramers $f^3$ system having the Kramers
doublet ground state. The fine electronic structure of the
U$^{3+}$ ion consists of five Kramers doublets as the effect of
charge interactions. These doublets are split in the
antiferromagnetic state, i.e. below T$_N$ of 14 K in case of
UPd$_2$Al$_3$. We have found [6] that the localized excitations
revealed by inelastic-neutron-scattering experiments of Krimmel et
al. [9] are in reasonable agreement with our earlier predictions
for the U$^{3+}$ state.

The motivation for this paper is to provide further arguments for
the existence of the localized $\emph{f}$-electron states of the
U$^{3+}$ configuration and to provide consistent explanation for
properties of NpPd$_2$Al$_3$. In the trivalent uranium state we
follow our very early explanation of the overall temperature
dependence of the specific heat with localized states associated
with the U$^{3+}$ configuration [5]. Our solution is along the
line of our consequent long-lasting approach to transition-metal
compounds underlying the importance of the crystal field,
including heavy-fermion compounds. One of us (RJR) bears with the
proudness a nick-name of Mister Crystal Field given by V.
Sechovsky in 1991, i.e. in years, when the use of the
crystal-field was largely discriminated within magnetic and
heavy-fermion community. It is worth to remind that 10 years ago
the crystal-field topic was simply expelled from the strongly
correlated electron systems conference.

\section{Theoretical idea}

The inelastic-neutron-scattering experiments [9] have revealed at
25 K the existence of the crystal-field excitations with energies
of 7 and 23.4 meV. This experiment at 150 K has revealed further
excitations at 3 and 14 meV at the energy-loss side and at 7 meV
at the energy-gain side. We have interpreted these excitations as
related to the energy level scheme: 0, 7 (81 K), 10 (116 K) and
23.4 (271 K) and we have ascribed this scheme to the $f^{3}$
(U$^{3+}$) scheme. There the specific-heat analysis [5] has
yielded the energy levels at 0, 100, 150, 350 and 600 K. All of
them are Kramers doublets. We have noticed in Ref. [6] that the
INS scheme is similar to that derived from specific heat analysis
provided that the energy scheme is reduced by about 80 percent. We
have derived a new set of CEF parameters of the hexagonal
symmetry: B$_2^0$=+5.3 K, B$_4^0$=+40 mK, B$_6^0$=-0.02 mK and
B$_6^6$=-26 mK.

For the description of electronic and magnetic properties we have
applied a single-ion like Hamiltonian for the ground multiplet
J=9/2 [6]:

\begin{center}
$H=H_{CF}+H_{f-f}=\sum \sum B_n^mO_n^m+ng^2\mu _B^2\left(
-J\left\langle J\right\rangle +\frac 12\left\langle J\right\rangle
^2\right) $
\end{center}

The first term is the crystal-field Hamiltonian.The second term
takes into account the intersite spin-dependent interactions that
produce e.g. the magnetic order below T$_N$ what is seen in Fig. 1
as the appearance of the splitting of the Kramers doublet.

\section{Results and discussion}

The derived electronic structure is shown in Fig. 1. The derived
parameters yield states at 81 K, 120 K, 270 K and 460 K. These
energies are in very good agreement with the INS data. The highest
state was not observed in INS experiment. This scheme yields the
specific heat that has very similar shape as in Ref. 5, a maximum
is only shifted to lower temperatures by about 10 K, but we think
that it is within the experimental accuracy. Such good description
of both thermodynamical property (specific heat) and spectroscopic
excitations we take as very strong confirmation of the $f^{3}$
(U$^{3+}$) state in UPd$_2$Al$_{3}$.
\begin{figure}[ht]
\vspace {-0.4cm}
\begin{center}
\includegraphics[width = 12.5 cm]{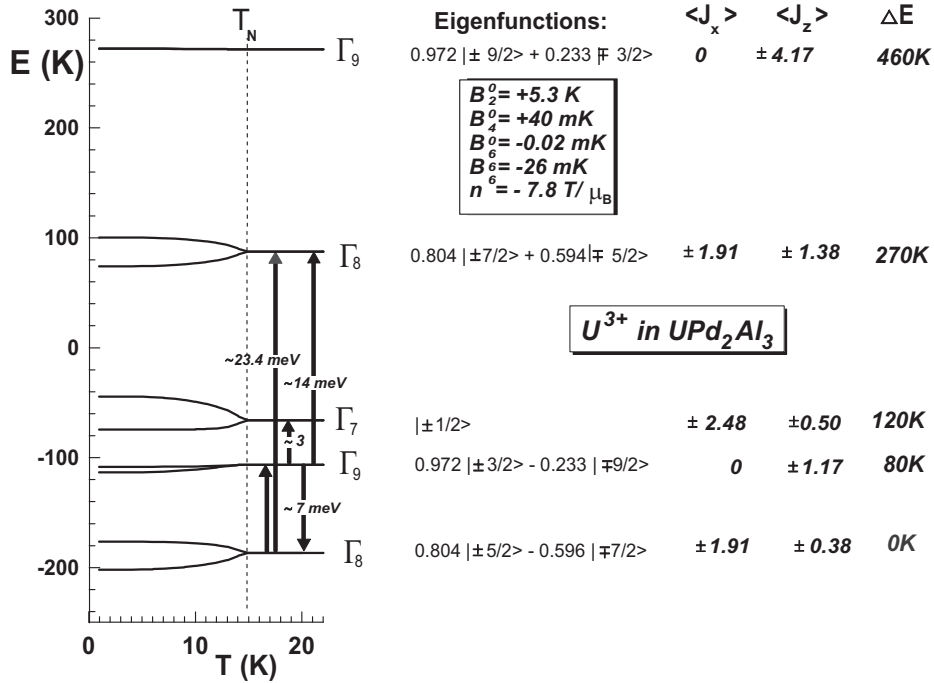}
\end{center}
\vspace {-0.8cm}\caption{Energy level scheme of the U$^{3+}$ ion
in UPd$_2$Al$_3$. Arrows indicate transitions that we have
attributed to excitations revealed by inelastic-neutron-scattering
experiments [9].}
\end{figure}
We are convinced that the derived U$^{3+}$ state in UPd$_2$Al$_3$
is consistent with the trivalent Np state in the isostructural
NpPd$_2$Al$_3$ derived by the Mossbauer spectroscopy [10]. In
order to exploit further the CEF theory we recalculate CEF
parameters from UPd$_{2}$Al$_{3}$ to NpPd$_2$Al$_3$ using
single-ion theory, i.e. making use of the Stevens coefficients. We
found the CEF parameters for Np$^{3+}$: B$_2^0$=-5.7 K,
B$_4^0$=-50 mK, B$_6^0$=+0.027 mK and B$_6^6$=-36 mK. The
resulting energy level scheme is shown in Fig. 2.
\begin{figure}[ht]
\vspace {-0.5cm}
\begin{center}
\includegraphics[angle=270, width = 13.5 cm]{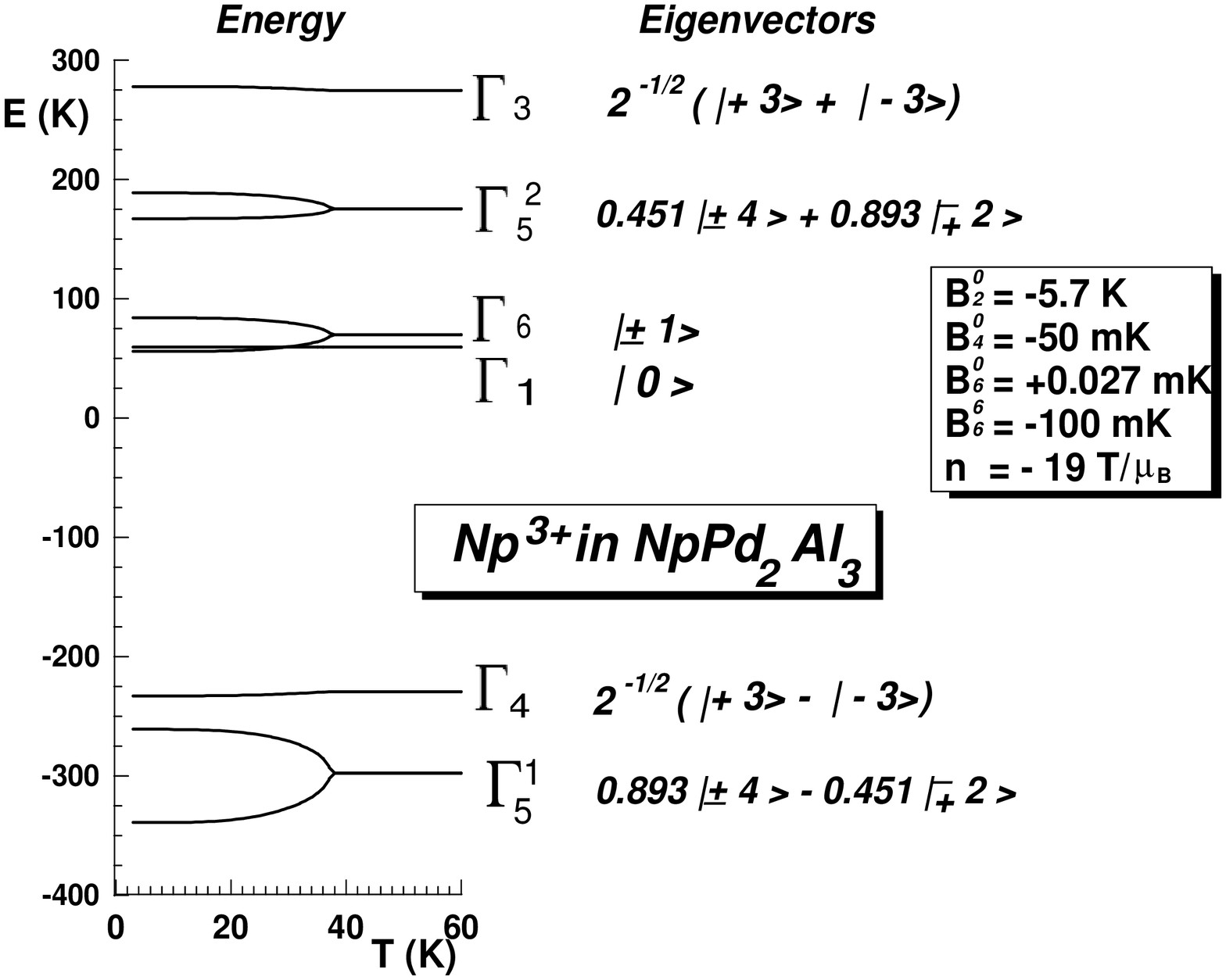}
\end{center}
\vspace {-0.8cm}\caption{The energy level scheme of the Np$^{3+}$
ion in NpPd$_2$Al$_3$.}
\end{figure}
These parameters yield the magnetic moment direction of the
Np$^{3+}$ ion to lie along the hexagonal c axis, i.e. the
perpendicular to the uranium moment. Exactly as is observed.
However, the moment value amounts to 2.2 $\mu_{B}$ that is larger
than a Mossbauer estimation of 1.6 $\mu_{B}$ [10]. We manage to
reduce the value of the magnetic moment, not loosing the c-axis
direction by increasing the value of the B$_6^6$=-36 mK parameter
to B$_6^6$=-100 mK. With this value we get magnetic moment of 1.6
$\mu_{B}$ in the paramagnetic state and 1.79 $\mu_{B}$ in the
ordered state, Fig. 3.
\begin{figure}[ht]
\begin{center}
\includegraphics[angle=270, width = 12 cm]{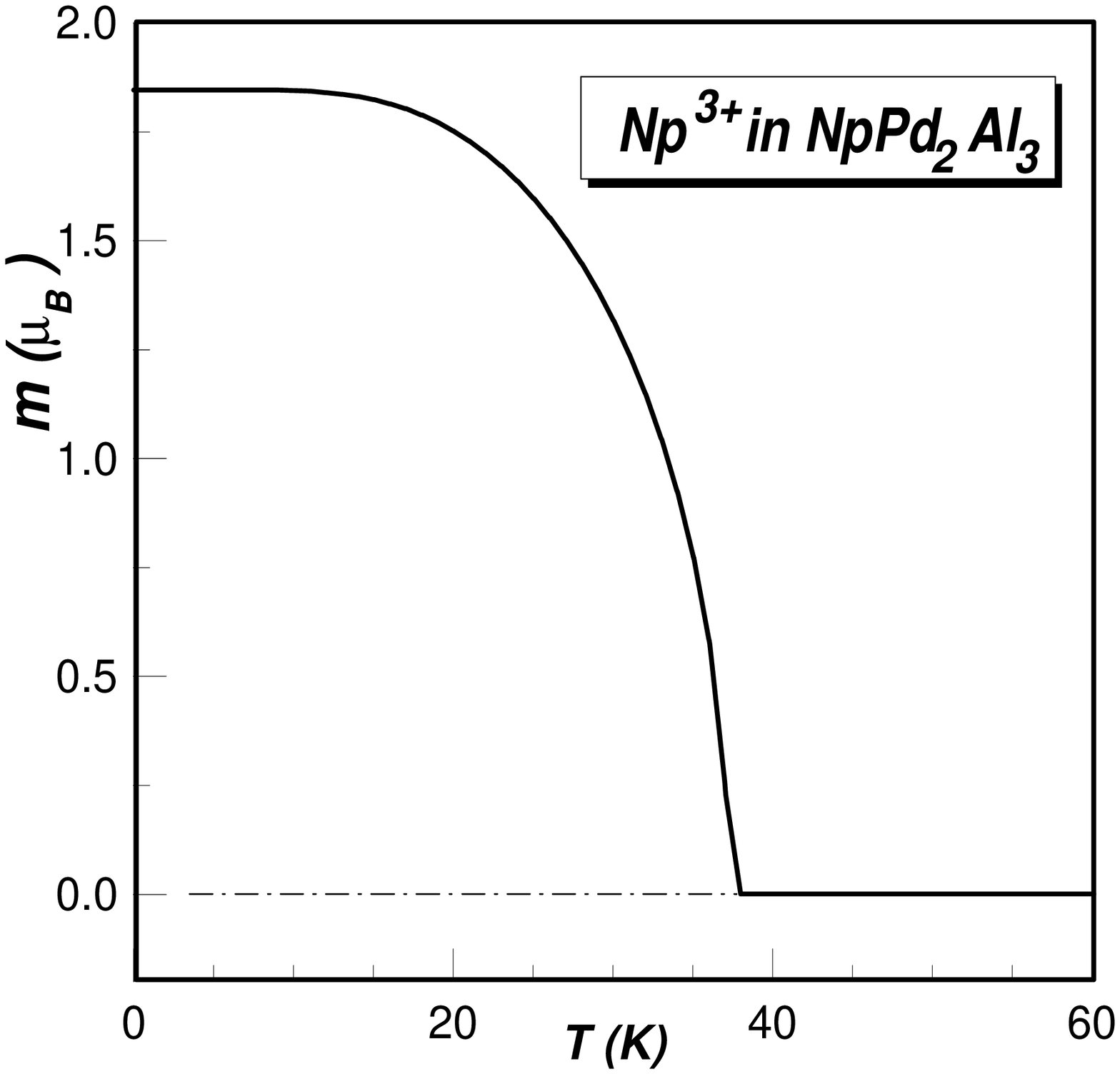}
\end{center}
\vspace {-0.8cm} \caption{Calculated temperature dependence of the
Np$^{3+}$ ion in NpPd$_2$Al$_3$.}
\end{figure}
We consider the shown parameters to be quite close to the real
situation, however, more experiments are awaited, in particular
the observation of the highest state in the
inelastic-neutron-scattering experiment.

\section{Conclusions}

It is pointed out that the experimentally-observed crystal-field
excitations in the heavy-fermion superconductor UPd$_2$Al$_3$ are
consistent with the U$^{3+}$ state. It makes UPd$_2$Al$_3$ to be
the second, apart UPd$_3$, uranium compound in which CEF states
have been unambiguously revealed. The existence of the localized
$f$ states in heavy-fermion superconductor UPd$_2$Al$_3$ is taken
as a strong support for the validity of the crystal-field model
for the h-f phenomena in which the crystal-field realization of
the Kramers-doublet ground state is the basic ingredient.

The derived U$^{3+}$ state in UPd$_2$Al$_3$ is consistent with the
trivalent Np state in the isostructural NpPd$_2$Al$_3$ derived by
the Mossbauer spectroscopy. Our set of CEF for UPd$_{2}$Al$_{3}$
yields the magnetic moment direction of the Np$^{3+}$ to lie along
the hexagonal c axis, i.e. perpendicularly to the uranium moment,
exactly as is experimentally observed.

Description of so many different properties of UPd$_2$Al$_3$ and
successful prediction of the magnetic properties of NpPd$_2$Al$_3$
we take as being not coincidence only but as indication of large
physical adequacy of the developed by us the Quantum Atomistic
Solid State Theory (QUASST) [11,12]. QUASST points out the
necessity of discussion of electronic and magnetic properties on
the atomic scale where the local symmetry, crystal field and
strong electron correlations, of intra-atomic and inter-site
origin, plays the fundamental role in formation of the low-energy
discrete energy spectrum. The experiment by Krimmel \emph{et al.}
[9] we regard as a mile stone in the experimental confirmation of
the QUASST theory for intermetallic, where localized and itinerant
(conduction) electrons coexist.

\end{document}